\documentclass[review]{elsarticle}

\usepackage{}
\journal{Journal of \LaTeX\ Templates}

\begin{document}
\begin{frontmatter}

\title{Gapped  vegetation patterns:  crown/root allometry and  snaking bifurcation} 
%\tnotetext[mytitlenote]{Fully documented
%templates are available in the elsarticle package on
%\href{http://www.ctan.org/tex-archive/macros/latex/contrib/elsarticle}{
%CTAN}.}

\author{Jaime Cisternas} 

\author{Daniel Escaff} 
\address{Complex Systems Group, Facultad de Ingenier\'{\i}a y Ciencias
Aplicadas, Universidad de los Andes, 
Monse{\~n}or Alvaro del Portillo  12455, Las Condes, Santiago, Chile.}

\author{Marcel G. Clerc}
\address{Departamento de F\'{\i}sica, Facultad de Ciencias F\'{\i}sicas y Matem\'aticas,
Universidad de Chile, Casilla 487-3, Santiago, Chile.}

\author{Ren\'e Lefever}

\author{Mustapha Tlidi\corref{mycorrespondingauthor}} \ead{mtlidi@ulb.ac.be}
\cortext[mycorrespondingauthor]{Corresponding author}
\address{Facult\'e des Sciences, Universit\'e Libre de Bruxelles (ULB), Campus Plaine, CP. 231, B-1050 Bruxelles, Belgium.}

\begin{abstract} 
Nonuniform spatial distributions of vegetation in scarce environments consist of either gaps, bands often called tiger bush 
or patches that can be either self-organized or spatially localized in space.  
When the level of aridity is increased, the uniform vegetation cover develops localized regions of lower biomass.
These spatial structures are generically called vegetation  gaps. 
They are embedded in a uniform vegetation cover. The spatial distribution of vegetation gaps can be 
either periodic or  randomly distributed. We investigate the combined influence of 
the  facilitative and the competitive nonlocal interactions between plants, 
and  the role of crow/root allometry,  on the formation of gapped vegetation patterns. 
We characterize first the 
formation of the periodic distribution of gaps by drawing their bifurcation diagram. We  then characterize localized and aperiodic distributions of vegetation gaps  in terms of their snaking bifurcation diagram. 
\end{abstract}

\begin{keyword} \texttt{elsarticle.cls}\sep \LaTeX\sep Elsevier \sep
template \MSC[2010] 00-01\sep  99-00 \end{keyword}

\end{frontmatter}

\section{Introduction}
When the level of the aridity increases, vegetation populations exhibit two-phase structures where high biomass areas are separated by sparsely covered or even bare ground. The spatial fragmentation of landscapes leading to the formation of vegetation patterns in semi- and arid-ecosystems is attributed to the symmetry-breaking instability that occurs even under strictly homogeneous and isotropic environmental conditions \cite{LL97}.   On-site measurements supported by theoretical modeling showed indeed that vegetation patterns are formed thanks to the facilitation and competition plant-to-plant interactions \cite{BCLDL08,LBCL09,couteron2014plant,tlidi2018observation}. 
The periodic vegetation patterns that emerge at the onset of this instability are characterized by a well-defined wavelength. Generally speaking, symmetry-breaking instability requires two opposite feedbacks that act on different spatial scales. The positive feedback consists of the facilitative plant-to-plant interaction such as shadow and shelter effect. This feedback operates on a small spatial scale comparable to the size of the plant crown and tends to improve the water budget in the soil  \cite{callaway1995positive},  
which thus favors the increase of vegetation biomass in arid-zones. 
The negative feedback by competitive plant-to-plant interaction for resources such as water and nutrients which on the contrary operates on a longer space scale corresponding to the plant lateral root length.   The roots of a given plant tend to deprive its neighbors' resources such as water
\cite{callaway1997competition}. The balance of competition and facilitation interactions within 
plant communities allow for the stabilization of vegetation patterns.
This symmetry-breaking instability is characterized by an intrinsic wavelength that is solely determined by dynamical parameters such as the structural parameter (the ratio between the size of the crown and the rhizosphere) and the vegetation community cooperatively parameter or other internal effects \cite{turing1952chemical,prigogine1968symmetry}.   The morphologies of these states follow the generic sequence  
gaps $\iff$ bands or labyrinth $\iff$ spots
as the level of the aridity is increased  \cite{LT99,QUA}. This generic scenario has been recovered for other mathematical models that incorporate water transport    by below ground  and/or above ground run-off   \cite{von2001diversity,Rietkerk2008169,PhysRevLett.93.098105,Gowda}.

\begin{figure}
\begin{center}
\includegraphics[width = 12cm]{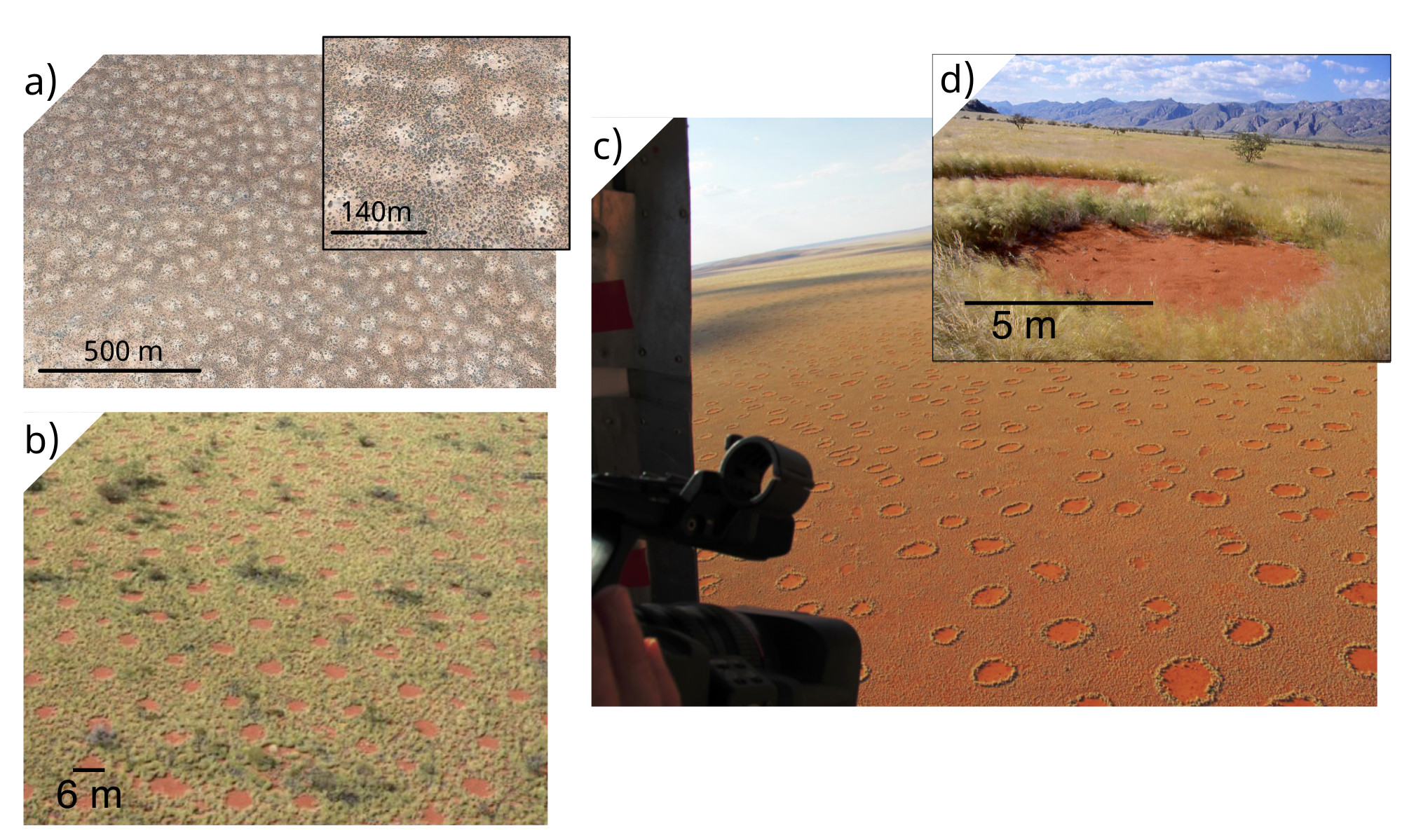}
\end{center}
\caption{Field observations of gapped vegetation patterns. 
(a) North western Africa (Kenya, $0º42'16.61'' $N, $40º22'19.06'' $E ). 
(b) North west Australia (c,d) Pro-Namibia zone of the west coast of Southern Africa 
(c) Aerial photography: courtesy of Norbert Juergen. (d) Photography: courtesy of Johnny Vergeer .} \label{fig00} \end{figure}

Vegetation patterns are not necessarily periodic. They can be aperiodic and localized in space in the form of more or less circular patches surrounded by a bare soil \cite{PhysRevE.66.010901,Rietkerk_Science_04}. Localized vegetation structures is a patterning phenomenon that occurs under the same condition as symmetry-breaking instability.  However,  for a moderate level of aridity, they tend
to spread and to invade the whole space available in a given landscape. This bifurcation is referred to as curvature instability that deforms the circular shape of localized patches and provokes a self-replication phenomenon that can take place even in strictly isotropic environmental conditions  \cite{bordeu2016self,bordeu2015localized,tlidi2018extended}.  This curvature instability may lead to the formation of another type of morphologies, such as arcs and spiral 
like vegetation patterns  \cite{tlidi2018observation}.

We investigate the formation of gapped vegetation patterns under a strictly homogeneous environmental conditions.  
They consist either periodic or aperiodic distribution of spots of bare soil  embedded in a uniform vegetation cover (see Fig.~1).  
By taking into account the effect of the crown/root allometry together with the facilitative and the competitive plant-to-plant interactions, we show that the homogeneous cover exhibits a symmetry-breaking instability leading the formation of spatially periodic distribution of gaps. 
We characterize first the formation of periodic gaps by drawing their bifurcation diagram.
Then, we identify the range of the parameter where landscapes exhibit stable localized gaps.
These structures are characterized by an exponentially decaying oscillatory tails, which stabilize a large number of gaps and clusters of them  \cite{TLV08}. We establish a snaking bifurcation diagram associated with localized gaps. 
From the theory of dynamical systems point of view, the bifurcation diagram associated with gaps localized structure consists of two snaking curves: one describes
gaps localized vegetation patterns within an odd number of gaps, the other corresponds to even a number of gaps. 
This type of gaps self-organization is often called homoclinic snaking bifurcation and have been intensively  
studied in the framework of the Swift-Hohenberg equation 
\cite{woods1999heteroclinic,coullet2000stable,clerc2005localized}, 
see also a review paper \cite{burke2007homoclinic} in the theme issue \cite{tlidi2007introduction}.  
A paradigmatic Swift-Hohenberg equation has been derived in the context of plant ecology 
in vicinity of the critical point associated with bistability by Lefever and collaborators \cite{LBCL09}.  
This model undergoes snaking bifurcations for both localized patches and gaps \cite{vladimirov2011relative,bel2012gradual}.
 
The paper is organized as follows: In section 2, 
we present the model and analyze the uniformly covered states. 
In section 3, we analyzed and characterize the formation of gapped vegetation patterns in one and two-dimensions. 
In section 4 we study localized gaps, and analyze them in terms of the homoclinic snaking bifurcation in one dimension.

\bigskip
\section{Interaction-redistribution model for the biomass evolution: nonlocal interactions and crown/roots allometry}
During more than two decades, several models have been proposed to investigate the formation of  vegetation patterns and the associated self-organization phenomenon. 
These approaches  can be classified into three categories.  The first is the generic interaction-redistribution model based on the relationship between the structure of individual plants and the facilitation-competition plant-to-plant interactions existing within plant communities. This modeling considers a single biomass variable \cite{LL97}.   The second is based on reaction-diffusion type of modeling. 
This approach focuses on the  influence of water transport by below ground diffusion and/or above ground run-off   \cite{klausmeier1999regular,hillerislambers2001vegetation,von2001diversity,Rietkerk2008169,PhysRevLett.93.098105}.
The third modeling approach is based on the stochastic processes  that take into account the role of environmental randomness as a mechanism  of noise induced symmetry-breaking transitions \cite{d2006patterns,ridolfi2011noise}. 
In what follows, let us consider the first modeling approach, the interaction-redistribution model \cite{LT12},
\begin{equation} \partial_t b = F_f\left(b + d
F_d\right)\left(1 - F_c\right) - \mu b,\label{MODEL} 
\end{equation}
where $b=b({\bf r},t) $ is the biomass density at the position ${\bf r}=(x,y)$ and the time $t$.
 The functions  $F_f$, $F_c$, and $F_d$ account for facilitation,  competition, and seed dispersion mechanisms of the plant-to-plant feedbacks, respectively. The parameter $\mu$ models the \emph{aridity}, that
is, the plant mortality, which is mostly attributed to adverse environmental conditions.  The explicit forms of the nonlocal functions $F_{f,c,d}$, are
 \begin{eqnarray*} F_f= \exp{\left( \frac{1}{C_1} \int e^{ -
\frac{\left|{\bf r}^{\prime} \right|}{L_1} }
b({\bf r}+{\bf r}^{\prime},t)\ d {\bf r}^{\prime} \right)},& & 
F_c= \frac{\int e^{ -
\frac{|{\bf r}^{\prime}|}{L_2}}
b({\bf r}+{\bf r}^{\prime},t)\ d{\bf r}^{\prime} }{\int e^{ -
\frac{|{\bf r}^{\prime}|}{L_2}}\ d
{\bf r}^{\prime}}, \\   \qquad F_d &=& \frac{1}{C_d} \int e^{-
\frac{{\bf r}^{\prime 2}}{L_d^2} } b({\bf r}^{\prime}+{\bf r},t)\ d {\bf r}^{\prime}.
\end{eqnarray*}
Normalization factors are $C_1=2 L_a$, and $C_d=\sqrt{\pi}L_d$ for one-dimensional systems, and 
$C_1=2 \pi L_a^2,$ and  $C_d=\pi L_d^2$ for two-dimensional systems.
$L_a$ is the effective radius of the surface that occupies a mature plant. However, ecosystems not only  
comprise  mature plants but also classes of different ages.  Indeed, the age, the crown size, and root sphere are different from each plant. Mature  and bigger plants require a higher amount of water and nutrients. 
Young and smaller plants explore through their roots smaller territories for the uptake of water and nutrients. In other words, the allometric factor plays an important role in
phyto-societal behaviors that governs the range of facilitative and competitive interactions during development of plants.
For a given spatial point, the effective ranges of the facilitative and the competitive interactions 
depend on the biomass density. The range of influence of young
plants (associated with lower biomass) must be smaller than the influence range of the older ones (associated with higher biomass). The allometric factor is a statistical constant that has been established to be $p=1/3$ by measuring  the relative size of the above-ground (crown) and below-ground (rhizosphere) structures for the plant C. micranthum in Neger  \cite{BCLDL08,LBCL09}.
The effective ranges of interaction take the form \cite{PhysRevLett.93.098105,sheffer2007plants}
\begin{equation} L_{1,2}= L_{1,2}^0b^p
\end{equation} 
 where $p$ is the \emph{allometric factor} and $L_{1,2}^0$ are constants.
 In the absence of the allometry, i.e., $p=0$ the effective ranges of both facilitative and competitive 
 interactions are independent of the biomass density. However as we shall see, 
 when $p \neq 0$, the allometry modifies the position of the critical point associated with bistability and induces a new branch of low biomass \cite{LBCL09,LT12}. 
 
The homogeneous steady states, $b({\bf r},t) = b_0$, corresponding to the homogenous covers are solution  of 
Eq.~(\ref{MODEL}). The trivial solution is the bare state $b_0=0$, represents a territory totally devoid of vegetation. The barren state obviously exists for all values of the parameters. Homogeneous covers satisfy the following equation
 \begin{equation} \mu =
\exp\left(\Gamma^n b_{0}^{np+1}\right)\left(1 + d\right)\left(1 -
b_0\right),
\label{ROOT2} 
\end{equation}  where $\Gamma \equiv L_1/L_a$ and $n$ is the dimension of the 
system ($n=1$ corresponds to 1D and $n=2$ corresponds to 2D). 
The dimension of the system and the allometric factor impact the homogeneous steady state curves.  Depending on the parameter values, one can find monostable and bistable regimes, between these
spatially homogeneous states. The coordinates of this second-order critical point marking the onset of a hysteresis loop is obtained by satisfying simultaneously the conditions
\begin{equation}
\frac{\partial \mu}{\partial b_0} = 0 ~~~~ \mbox{and} ~~~~  
\frac{\partial^2\mu}{\partial b_{0}^{2}} = 0,
\label{MONO-BI}
\end{equation} 
with $\mu$ satisfies Eq.~(\ref{ROOT2}).  
The two conditions Eq.~(\ref{MONO-BI}) allow to estimate the coordinate of the critical point associated with bistability  
$(\Gamma_c,\mu_c, b_c)$.

\begin{figure}
\includegraphics[width = 5in]{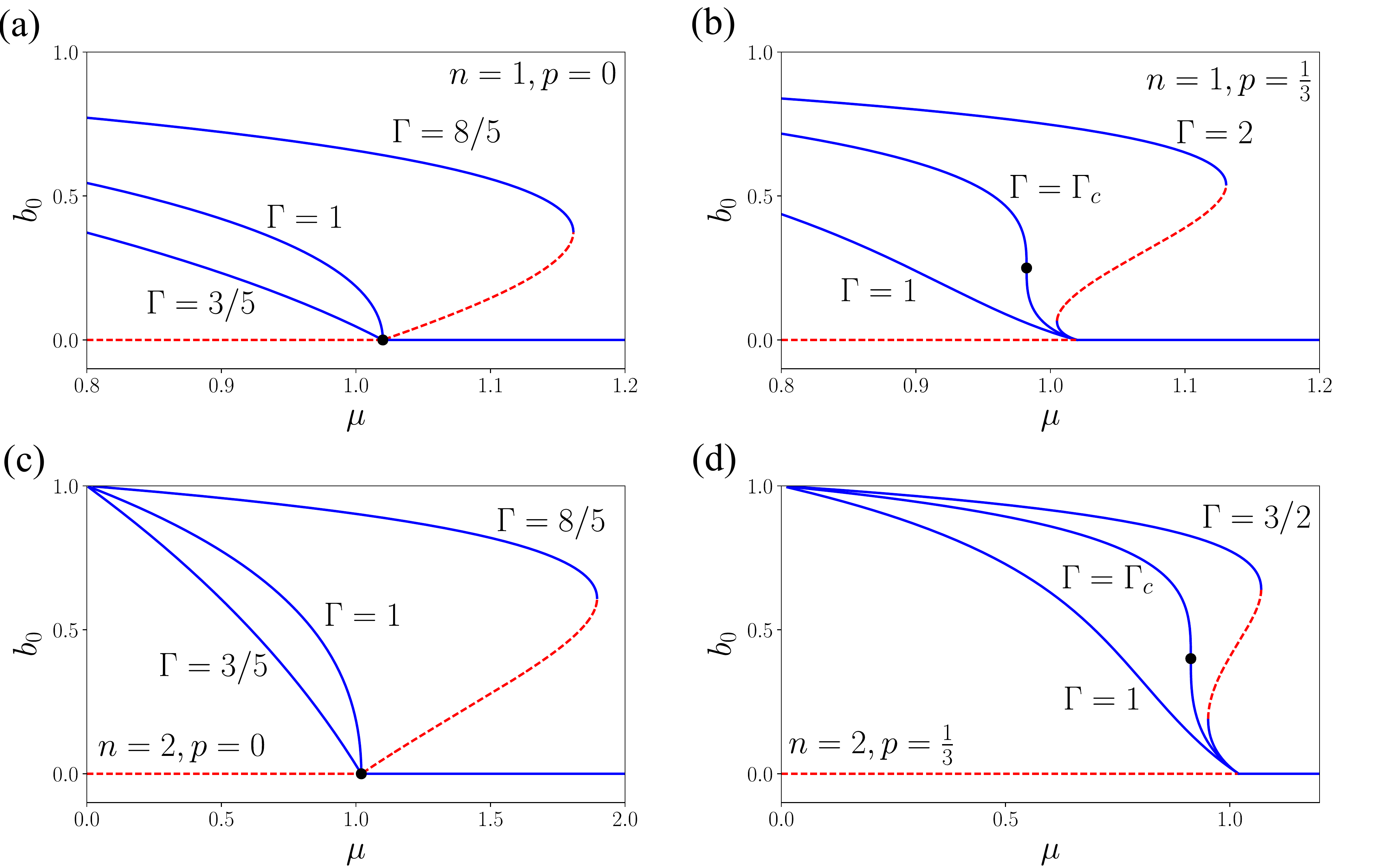}
\caption{Homogeneous steady states solutions of
Eq.~(\ref{ROOT2}), for $d=0.02$. The barren state $b_0=0$  is unstable for $\mu<1+d$; stable for $\mu>1+d$. Depending on the
allometric exponent $p$, the spatial dimensions $n$, and the ratio
$\Gamma$, the branch of vegetated solutions may or may not have an
unstable part and a second section of stability. (left top panel) 1-d, no allometric
effect $p=0$; (left bottom panel) 2-d, no allometric effect $p=0$; (right top panel) 1-d, allometric
effect $p=1/3$; and (right bottom panel) 2-d, allometric effect $p=1/3$. Stability to
sinusoidal perturbations is analyzed in another section.
}
\label{fig02} \end{figure}

In the absence of the allometry $p=0$, i.e., when all plants are mature, two regimes must be distinguished according to the value of the  
$\Gamma = L_1/L_a$ parameter. If $\Gamma<1$ (monostable regime), the uniformly vegetated  cover exists only in the range  
$0<  \mu < 1$. In this parameter range, the biomass decreases monotonously with aridity and vanishes at $\mu = 1+d$. For larger aridity levels, i.e., 
$\mu >1+d$ only the barren state exists as shown in Fig.~\ref{fig02}(a). If $\Gamma > 1$ (bistable regime), the state of the uniformly vegetated cover  extends up to the tipping point (saddle-node bifurcation) as shown in Fig.~\ref{fig02}(a,b).  This means that when increasing the facilitative interaction range,  vegetation community can survive while individual plants can not. This situation corresponds to the vegetation systems presented in Fig.~\ref{fig02}(a) in a one-dimensional system and in Fig. \ref{fig02}(b) for two-dimensional settings. If $\Gamma  =1$, the system reaches the  second-order critical point marking the onset of a hysteresis loop. 
The coordinate of this critical point is  $\mu_c = 1+d$ and $b_c=0$.  In this case, spatial oscillations around  $b_c=0$ are physically excluded.
When, however, we take into account the allometry, the coordinate of the critical point associated with  the
nascent bistability occurs with a finite biomass ($b_c \neq 0$), as shown in Figs.~\ref{fig02}(c,d). 
Another important consequence of the allometry is that a new stable state with low biomass density is possible as shown in Fig.~\ref{fig02}(d).  
Therefore, there is a parameter range where the system exhibits tristability. Besides the stable barren state, the high and low biomass covers can coexist for the same values of system parameters. From  Figs. \ref{fig02}, we infer that all curves, in
the $\left(\mu, b_0\right)$-plane, coincide at  the same point $\left(\mu = 1+d, b_0
=0\right)$. The slope at this point  is explicitly given by 
$\partial\mu/\partial b_0 = -\left(1 + d\right)$,
which is independent of the other parameter values. Dashed curves correspond to unstable solutions, while
continuous curves denote solutions that are stable with respect to  spatially homogeneous perturbations. In the next section, we will perform the linear stability analysis of the homogeneous covers with respect to spatially inhomogeneous perturbations.

\begin{figure}
\includegraphics[width =5in]{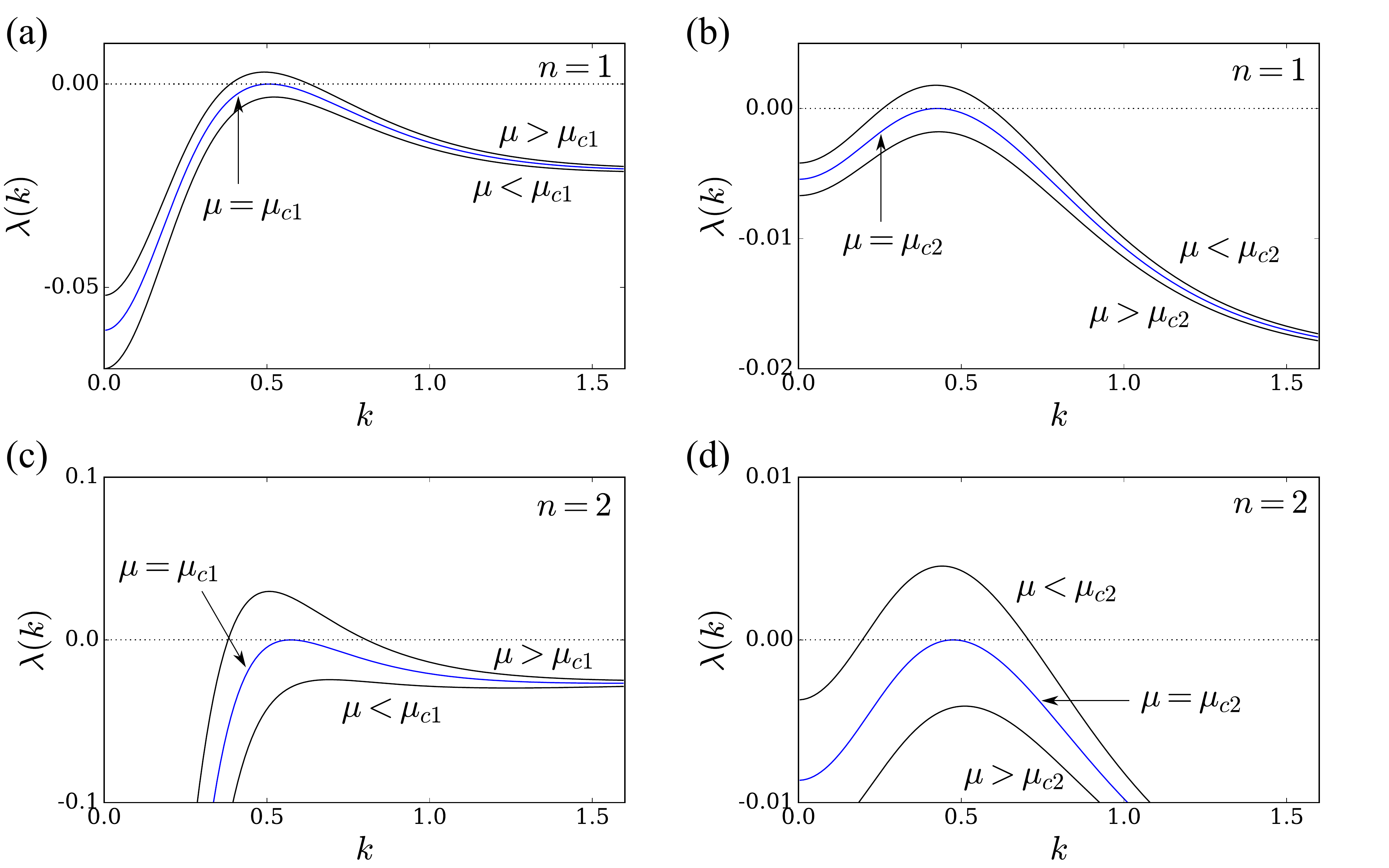}
\caption{Stability of uniformly vegetated state to the introduction of
small sinusoidal modulation, as parameter $\mu$ is varied, one ($n=1$) and two $(n=2)$
spatial dimensions. An allometric factor $p=1/3$ and a ratio
$\Gamma=5/8$ were used in all the figures, so there is a range defined
by two turning points where three vegetated homogeneous solutions exist.
(left panels) Along the upper solution there is a bifurcation point $\mu_{c1}$ at
which a critical mode $\exp(ikx)$ becomes unstable and its growth
exponent $\lambda(k)$ becomes positive. (right panels) Along the lower solution
there is also a bifurcation point $\mu_{c2}$ at which a mode becomes
unstable. Other parameters are $L_a=5/4$, $L_1 = 2$, $L_2 = 2.8125$, $L_d = 2.5$ and $d = 0.02$.
}
\label{fig03} \end{figure}

\section{Linear stability analysis and vegetation pattern formation}
\subsection{Linear stability analysis}
 We perform the linear stability analysis of the homogeneous steady state solutions of Eq.~(\ref{ROOT2}) 
 with respect to small  spatially inhomogeneous fluctuations around the homogeneous steady states; $b_0$; 
 of the form 
 \begin{equation} b({\bf r},t) = b_0 +
\delta b({\bf r},t),
\label{perturbation} 
\end{equation} 
with $\delta b({\bf r},t) \ll 1$. Replacing  Eq.~(\ref{perturbation}) into
Eq.~(\ref{MODEL}), and linearizing with respect to $\delta b({\bf r},t)$, one
obtains a linear equation of the form  $\partial_t \delta b({\bf r},t) = \mathcal{L}\delta b({\bf r},t)$. 
If all the eigenvalues of operator $\mathcal{L}$ have a negative real
part, the homogeneous state $b_0$ is stable, otherwise, it is unstable. Since
the system is invariant under spatial translations, the linear operator
$\mathcal{L}$ is diagonal in the Fourier basis 
$\mathcal{L}\exp\left(i {\bf k} \cdot {\bf r}\right) = \lambda(k) \exp\left(i {\bf k} \cdot {\bf r}\right)$.  
The function $\lambda(k)$, often called
the \emph{spectrum}, is always real for the approach based on the interaction-redistribution model Eq.~(\ref{MODEL})  and therefore
 time oscillations around the homogeneous covers are therefore excluded. The spectrum only depends on the
modulus of the wavevector ${\bf k} $. 
This is because the system is isotropic in both  $x$ and $y$ direction, 
and then there is no preferred direction in the plane  $(x,y)$. This spectrum can be computed analytically but the obtained expressions for both the threshold as well for the most unstable wavelength at the symmetry breaking instability are cumbersome. The plot of the spectrum as a function of the wavenumber is shown in Fig. \ref{fig03} for $p=1/3$.

It shows the spectrum for the upper
branches (left panels), that is, the higher density state, and the lower
branches (right panels), that is, the lower density state. These spectra
have been computed in one (top panels) and two (bottom panels)
dimensions and for $p=1/3$. As we see, from Fig. \ref{fig03},
there exist a range of aridity parameter,  $\mu_{c1}<\mu<\mu_{c2}$,
 where the homogeneous cover is unstable and giving rise to the formation of a periodic vegetation  pattern. 
 
\begin{figure}
\includegraphics[width =5in]{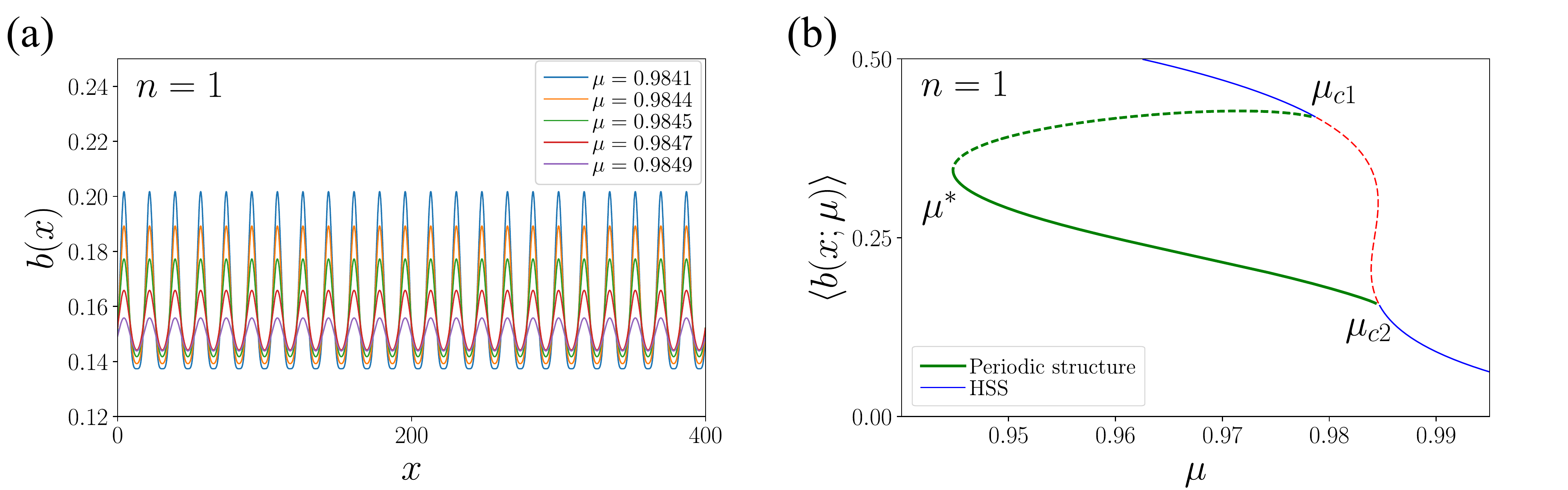}
\caption{One dimensional periodic structure. (a) Representation of
homogeneous solution and the spatially periodic solution that are born at the
two bifurcation points,
as computed with finite differences and pseudo-arclength continuation algorithm.
(b) Heterogeneous solutions for different values
of aridity $\mu$. As parameter $\mu$
decreases, the amplitude of the stable periodic profile grows. Other parameters are 
$L_a=5/4$, $L_1 = 2$, $L_2 = 2.8125$, $L_d = 2.5$, and $d = 0.02$.} \label{fig04}
\end{figure}

\subsection{Vegetation pattern formation}
To analyze the formation of vegetation patterns that are spontaneously triggered by  the symmetry-breaking instability, 
we first compute the spatially periodic structure in one spatial dimension $(n=1)$. For this purpose we use a continuation method. More precisely, we  numerically determine the branches of nonlinear solutions of  Eq.~(\ref{MODEL}). 
The stability analysis of these solutions is  performed with continuation software (AUTO), 
which is based on the pseudo-arclength method \cite{L14}. The results are summarized in Fig.~\ref{fig04},
where the spatially periodic profiles of the biomass are plotted for different values of the level of aridity $\mu$. 
From this figure, we see that as parameter $\mu$ is decreases, the biomass grows. 
The  bifurcation diagrams associated with these solutions are plotted as a function of the level of the aridity in  Fig. \ref{fig04}b.  
When increasing the level of aridity the  high biomass cover is stable in the range $0<\mu<\mu_{c1}$.
At $\mu=\mu_{c1}$, the  homogeneous high biomass cover becomes unstable with respect to 
spatially symmetry-breaking instability. 
From this bifurcation point $\mu=\mu_{c1}$, a periodic solution emerges spontaneously as shown in Fig.~\ref{fig04}b.  When increasing further the aridity, the homogeneous lower biomass state
stabilizes due to the second spatial symmetry instability at  $\mu=\mu_{c2}$ (see Fig.~\ref{fig04}b).
 The periodic solutions in the   vicinity of $\mu=\mu_{c2}$ appear supercritical with small  amplitude. However,  the bifurcation
 at $\mu=\mu_{c1}$ is subcritical. In this case, namely in the range $\mu^*<\mu<\mu_{c1}$, 
 the system exhibits  a coexistence  between a periodic gapped patterns and 
the homogeneous  high biomass cover.  As we will see in the next section, the coexistence is prerequisite conditions for the formation 
 of localized gaps. For $\mu>\mu_{c2}$, the low biomass cover is stable until the system reaches the barren state at $\mu=1+d$.
For $\mu>1+d$, only the barren state is stable.

\begin{figure}
\includegraphics[width =5in]{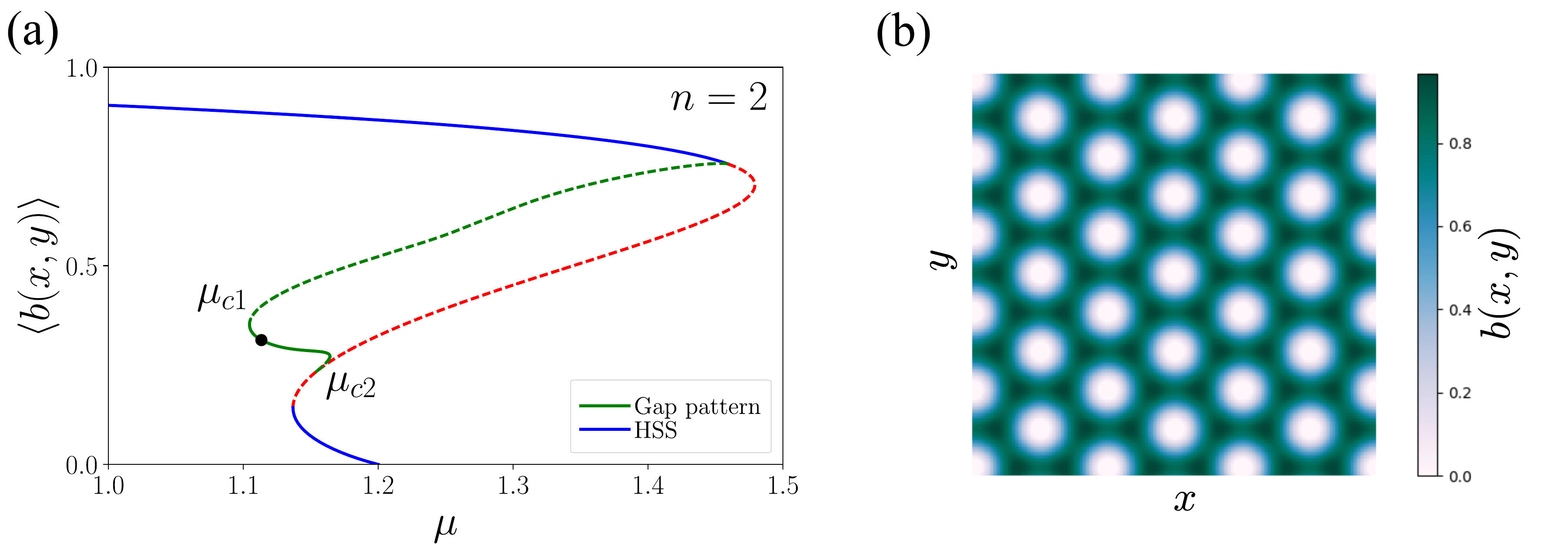}
\caption{(left panel) Branches of solutions with hexagonal symmetry
computed with a Fourier expansion and a pseudo-arclength continuation algorithm.
Parameter values: $p=1/3, L_a=5/4$, $L_1 = 2$, $L_2 = 2.8125$, $L_d = 2.5$ and $d = 0.02$.
(right panel) Stable hexagonal pattern of gaps for $\mu=1.11$ (indicated with black dot in left panel).} 
\label{fig04b} \end{figure}
In the two-dimensional system ($n=2$), when increasing the aridity parameter, the first vegetation pattern that appears is the periodic distribution of gaps forming a hexagonal structure as shown in Fig.~\ref{fig04b}. 
The bifurcation diagram associated with these two-dimensional solutions is plotted in  Fig. \ref{fig04b}. The results are obtained by using the continuation method in \cite{L14} that consist of assuming that the biomass is distributed in a periodic manner
and the wavevectors define a finite hexagonal lattice that is conjugated to the spatial lattice generated from the basic hexagon.

\begin{figure}[b]
\centering
\includegraphics[width =4.5in]{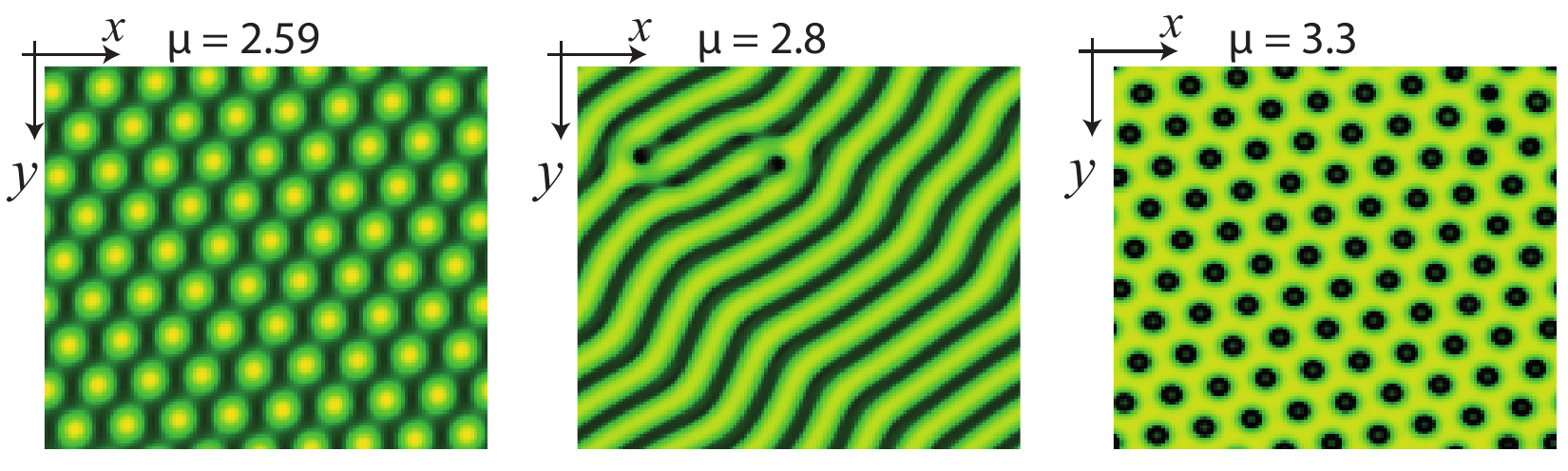}
\caption{Generic sequence of vegetation patterns obtained from model Eq.~(\ref{MODEL}). Aridity parameter is increases from left to right. 
other parameters are $d=0.5$, $p=0$, $L_a=5/4$, $L_1 = 2$, $L_2 = 2.8125$, and $L_d = 2.5$.} \label{fig04c}
\end{figure}

After having fully characterized the formation of  periodic distribution of gaps as a function of the aridity parameter, 
we recover the generic sequence  
gaps $\iff$ stripes $\iff$ spots as shown in Fig.~\ref{fig04c}. These ecological states in arid landscapes possess an overlap 
domain of stability as shown analytically in    
weak gradient approximation \cite{QUA}.  

\section{Localized gaps and snaking bifurcation}

Localized structures and localized patterns are a well 
documented phenomenon, concerning almost all fields of natural science including chemistry, biology, ecology, physics, 
fluid mechanics, and optics \cite{arecchi1999pattern,Akhmedievbook2008,TlidiPTRS-14,tlidi2016nonlinear,tlidi2018dissipative,Tlidi2018-part2}.  Localized vegetation patches and gaps  belong to the class of stationary and spatially localized patterns. 
They consist either of localized
vegetation patches distributed on bare soil \cite{PhysRevE.66.010901,meron2004vegetation} or, on the contrary,
of localized spots of bare soil embedded in an otherwise uniform
vegetation cover  \cite{TLV08}. Localized patches or localized bare soil spots
(often called fairy circles) can be either spatially independent, self-organized, or randomly distributed.

It is worth mentioning that fairy circles are striking examples attributed to this category of 
localized vegetation patterns \cite{van2004mysterious}. Although, the mechanisms leading to their formation are still subject of debate among the scientific community. 
They have been interpreted as the result of a pinning front between a uniform plant distribution 
 and a periodic hexagonal vegetation pattern  \cite{TLV08}. Recent investigations support this interpretation \cite{tschinkel2012life,cramer2013namibian,getzin2015adopting}. 
On the other hand, fairy circles formation may result from front dynamics connecting a bare state and uniform plant distributions \cite{fernandez2014strong,fernandez2013strong,escaff2015localized}. In all these works, the origin of fairy circle formation is intrinsic to the dynamics of the system. This means that the diameter of the fairy circle is  determined rather by the parameters of the system and not by external effects, such as the presence of social insects or anisotropy. However,  another theory based on external effects, such as termites or ants has been suggested \cite{becker2000fairy,picker2012ants,juergens2013biological,bonachela2015termite}. More recently, Tarnita  and collaborators have shown that a combination of intrinsic and extrinsic effects could explain the origin of fairy circles \cite{tarnita2017theoretical}.

\begin{figure}
\includegraphics[width =5in]{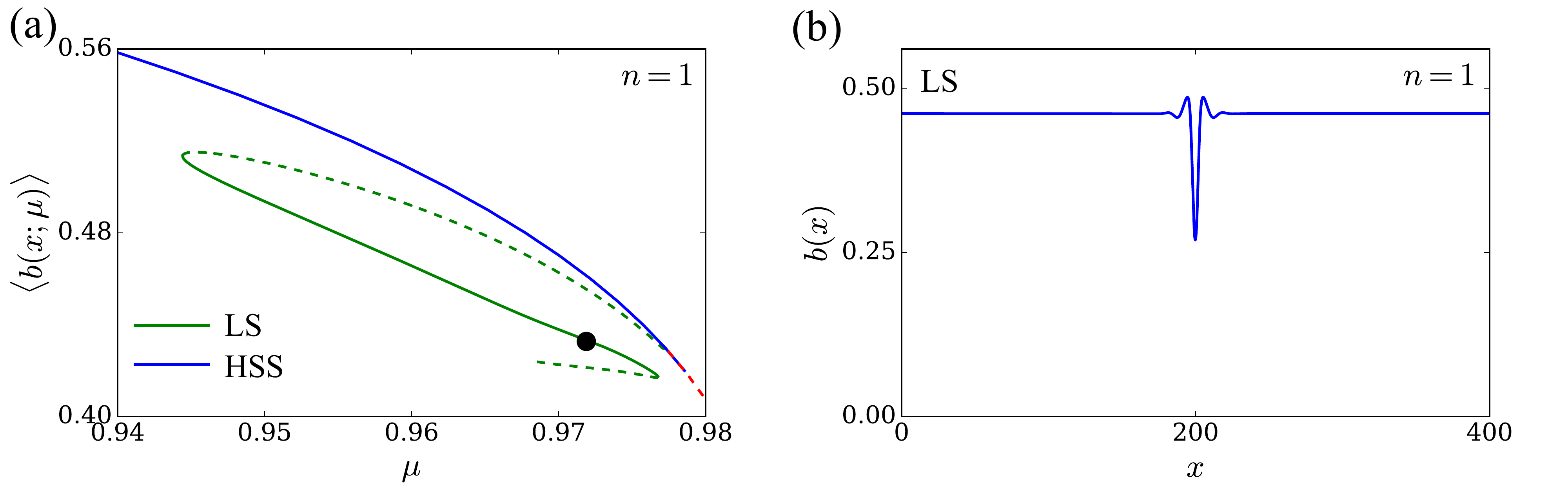}
\caption{(left panel) Branch of
localized solutions with odd number of gaps, born at the bifurcation
point of the upper homogeneous branch. The vertical coordinate is the
norm of the solution with respect to the homogeneous solution for that
value of $\mu$. (right panel) Localized solution with a single gap. This solution
is stable and corresponds to $\mu=0.972$ (black dot in left panel). Other parameters are $L_a=5/4$, $L_1 = 2$, $L_2 = 2.8125$, 
$L_d = 2.5$, and $d = 0.02$.} \label{fig05} \end{figure}

\begin{figure}
\includegraphics[width =5in]{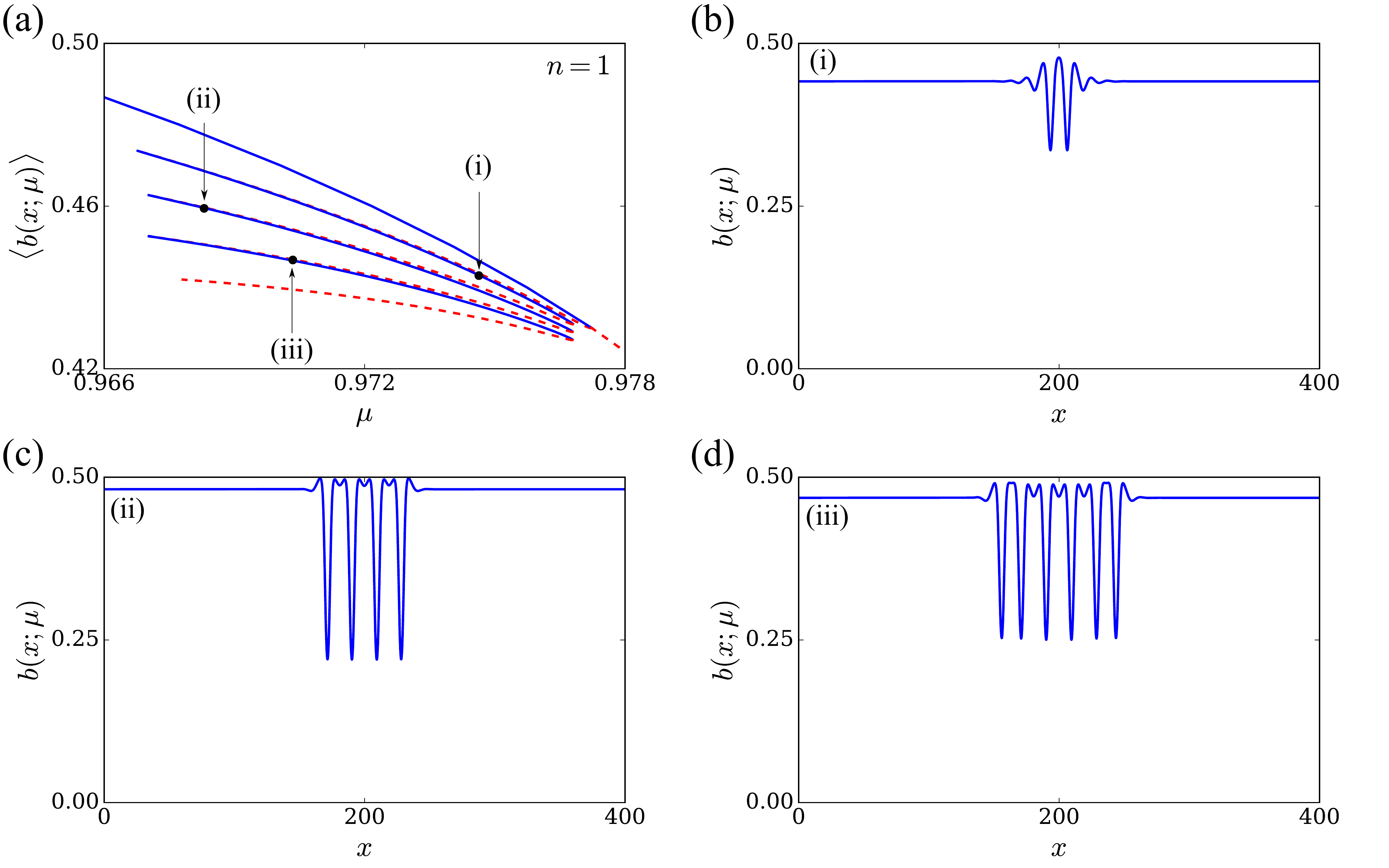}
\caption{(left panel) Branch of localized solutions with even number of gaps,
born at the bifurcation point of the upper homogeneous branch. The
vertical coordinate is the norm of the solution with respect to the
homogeneous solution for that value of $\mu$. (other three panels) Localized
solutions with two, three and four gaps. These solutions are stable and
correspond to $\mu=0.9754$, $\mu=0.9673$, and $\mu=0.9704$
(dots in left panel).
Other parameters are $L_a=5/4$, $L_1 = 2$, $L_2 = 2.8125$, $L_d = 2.5$, 
and $d = 0.02$.} \label{fig06} \end{figure}

%A minimum condition for the existence of such localized solutions is the
%coexistence of two homogeneus vegetated states as in
%Fig.~\ref{fig02}(b,c). 
The bifurcation theory of localized structures developed for
other extended models indicates that they are born at the same critical
point as vegetation gapped periodic patterns, and they are unstable, gaining stability
only after a fold. For instance a localized solution born at the
bifurcation point of the upper branch will look like a small region of
smaller density surrounded by a uniform region of density equal to the
upper branch for that value of the aridity, essentially a gap. A
localized gap pattern  born at the lower branch will look like a bump surrounding by high biomass 
density (see Fig.~\ref{fig05}).

Subsequent folds of the branch of localized solutions make them switch
stability and add extra gaps or bumps. The complete bifurcation diagram,
plotted using $\mu$ and the spatial average of $b(x)$, shows a
complicated snaking diagram with an infinite number of folds as the
patterns becomes more and more similar to the periodic solution.
Note  that there are two basic branches: one
with an odd number of gaps, and other with an even number of gaps. 
We draw only the branch corresponding to odd number of gaps.
We find numerically the first few folds of these two branches.
These branches are displayed in Fig.~\ref{fig05}  for single gap and Fig.~\ref{fig06} for clusters of odd number of gaps.

%In two spatial dimensions the previous picture becomes more complex as
%there is more than a single branch of localized solutions: now there are
%many ways to add bumps or gaps to the basic solution.....

%\begin{figure} \bigskip \caption{Localized solution in 2-d for $p=0$.
%Hexagonal pattern and its associated branch?} \label{fig07} \end{figure}

\bigskip

\section{Conclusions}

By using the generic interaction-redistribution model Eq.(\ref{MODEL}), 
we have analyzed ecological state transitions in stressed landscapes. We have shown that the 
allometry affects the homogeneous covers by : (i) shifting the critical point associated with bistability to a finite biomass density and (ii) by inducing a new branch of stable low biomass cover. We have shown that gaps are the first self-organized structure that appears when the level of the aridity is increased. We have attributed their formation to facilitative and competitive interactions between individual plants. We have recovered the generic sequence gaps $\iff$ bands or labyrinth $\iff$ spots.  We have established the  two-dimensional bifurcation diagram associated with periodic gapped vegetation patterns. To  perform the analysis,  we have used the  continuation method 
based on the pseudo-arclength technique.
Furthermore, we have fully analyzed the formation of localized gaps, and we have shown that they undergo a homoclinic snaking bifurcation. 

\section*{Funding}

This work was supported by a FONDECYT-Chile grant [1170669].  MT received support from 
the Fonds National de la Recherche Scientifique (Belgium). 

\end{document}